\documentclass[12pt]{iopart}

\usepackage{graphicx}
\usepackage{paralist}
\usepackage{cite}

\begin{document}

\title{Scanning-gate-induced effects and spatial mapping of a cavity}

\author{R Steinacher, AA Kozikov, C R\"ossler, C Reichl, W Wegscheider, T Ihn, and K Ensslin}

\address{Solid State Physics Laboratory, ETH Zurich, 8093 Zurich, Switzerland}

\ead{richard.steinacher@phys.ethz.ch}

\begin{abstract}

Tailored electrostatic potentials are at the heart of semiconductor nanostructures. We present measurements of size and screening effects of the tip-induced potential in scanning gate microscopy on a two-dimensional electron gas. First, we show methods on how to estimate the size of the tip-induced potential. Second, a ballistic cavity is studied as a function of the bias-voltage of the metallic top gates and probed with the tip-induced potential. It is shown how the potential of the cavity changes by tuning the system to a regime where conductance quantization in the constrictions formed by the tip and the top gates occurs. This conductance quantization leads to a unprecedented rich fringe pattern over the entire structure. Third, the effect of electrostatic screening of the metallic top gates is discussed.

\end{abstract}

\maketitle

\section{Introduction}

Scanning gate microscopy (SGM) is a powerful method to investigate local transport properties of electronic nanostructures. Typically the biased tip of a scanning force microscope is used to locally deplete (in the case of AlGaAs heterostructures, \cite{Eriksson1996, topinka2000, topinka2001, LeRoy2002, Jura2007, Jura2009, Jura2010, Kozikov2013imaging, Kozikov2013, Kozikov2014, brun2014wigner, Pascher2013, Paradiso2012, crook2003imaging}) or change (in the case of graphene, \cite{Berezovsky2010a, schnez2010, Pascher2012, Garcia2013}) the carrier density below the tip. The conductance is monitored as a function of tip position resulting in so-called scanning gate images. Important ingredients for the interpretation of such images are the shape and size of the tip-induced potential in the landscape of the electronic nanostructure. In the literature numbers for the size of the tip-induced potential at the Fermi energy vary between a few tens of nm and more than 1 µm depending on experimental setup and analysis procedure \cite{Eriksson1996, Pascher2012, Crook2000, Girard1993, Atlan1992, Pala2008, Pala2009, Chae2012, Martins2007, kivcin2005spatially, pioda2007discrete, Gildemeister2007, Sellier2011}. Since many nanostructures, such as quantum point contacts (QPCs) and quantum dots, are formed by suitably biased top gates, the effective electronic landscape is a superposition of the gate-defined and tip-induced potential.

In this paper we describe four different and complimentary methods which allow us to determine the effective size of the tip-induced potential at the Fermi energy. The sample is a gate-defined ballistic cavity. With the additional tip-induced potential we measure the positions of the conductance plateaus formed by the tip-gate constrictions and how they shift as a function of tip position. Equipped with this knowledge we also analyze the effects of gate screening on the detailed positions of observed features in scanning gate images. Our methods give a better understanding of the details of the potential landscape in complex gate geometries. Beyond that they are useful for scanning gate microscopy and in agreement with calculations.

\section{Experimental setup}

The investigated 2DEG is formed in a molecular-beam-epitaxy-grown GaAs/AlGaAs heterostructure with a density of $1.5\times 10^{11}\,\mathrm{cm^{-2}}$ and a mobility of $3.8 \times 10^{6}\,\mathrm{cm^2}/\mathrm{Vs}$ at a temperature of 300 mK. It is buried $120\,\mathrm{nm}$ below the surface. The electrons have a Fermi wavelength of $65\,\mathrm{nm}$ and an elastic mean free path of about $50\,\mu$m.

\begin{figure}
	\centering
	\includegraphics[scale=1]{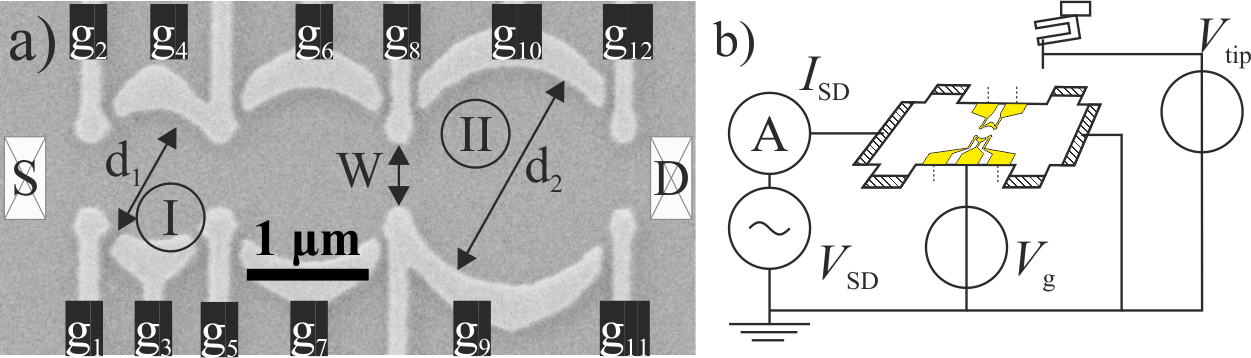} 
	\caption{a) Scanning electron micrograph of the sample used in most of the presented measurements. The bright parts correspond to the Ti/Au top gates $g_i$ placed on the GaAs surface (dark). b) Electric scheme of the measurement setup. The Hall bar is connected in a two-terminal configuration.}\label{fig1}
\end{figure}

The sample under study is fabricated by etching a conventional Hall bar. On top Au/Ti gates [see \fref{fig1}a)] are placed using electron beam lithography to define the two cavities in the following measurements. The segmented design is intended to give flexibility in forming cavities with different diameters ($d_1=1.0\,\mu$m for cavity I with gates $g_1-g_5$, $d_2= 1.5\,\mu$m for cavity II $g_8-g_{12}$). The lithographic width of the constrictions used as openings of cavity I (gates $g_1$ and $g_4$, $g_3$ and $g_4$, as seen in \fref{fig1}) is $0.62\,\mu$m. The constrictions used for cavity II ($g_8$ and $g_9$, and $g_{11}$ and $g_{12}$) are $W = 0.4\,\mu$m wide.

The experimental setup is a home-built AFM operated in a $^3$He cryostat \cite{Ihn2004} at a base temperature of $300\,$mK. A Pt/Ir wire, sharpened with chemical wet-etching and consecutive milling with a focussed ion beam is used as the tip. It is glued to a tuning fork sensor, which is controlled by a phase-locked loop \cite{rychen1999, rychen2000}.

The structure in the 2DEG is formed by applying negative voltages [$V_{g}$ in \fref{fig1}b)] to the top gates thereby decreasing the charge carrier density below the gates. The gate pinch-off is determined to be $-0.35\,$V. Biasing the tip ($V_\mathrm{tip} \approx -3\,..-8\,$V) 60 nm above the GaAs surface depletes the 2DEG underneath, and hence forms a movable gate. The transport measurements are carried out in a two-terminal configuration with a source-drain voltage ($V_\mathrm{SD}$) of $100\,\mu$V modulated at 27 Hz [\fref{fig1}b)]. The source-drain current ($I_{\mathrm{SD}}$) is measured by standard lock-in techniques.

\section{Tip depletion size in the 2DEG}

Information on the tip-induced potential is needed in order to interpret SGM results. In the following we show four methods which allow us to estimate the radius $R_\mathrm{tip}$ of the tip-depleted region in the plane of the electron gas.

\begin{figure}[h]
	\centering
	\includegraphics[scale=1]{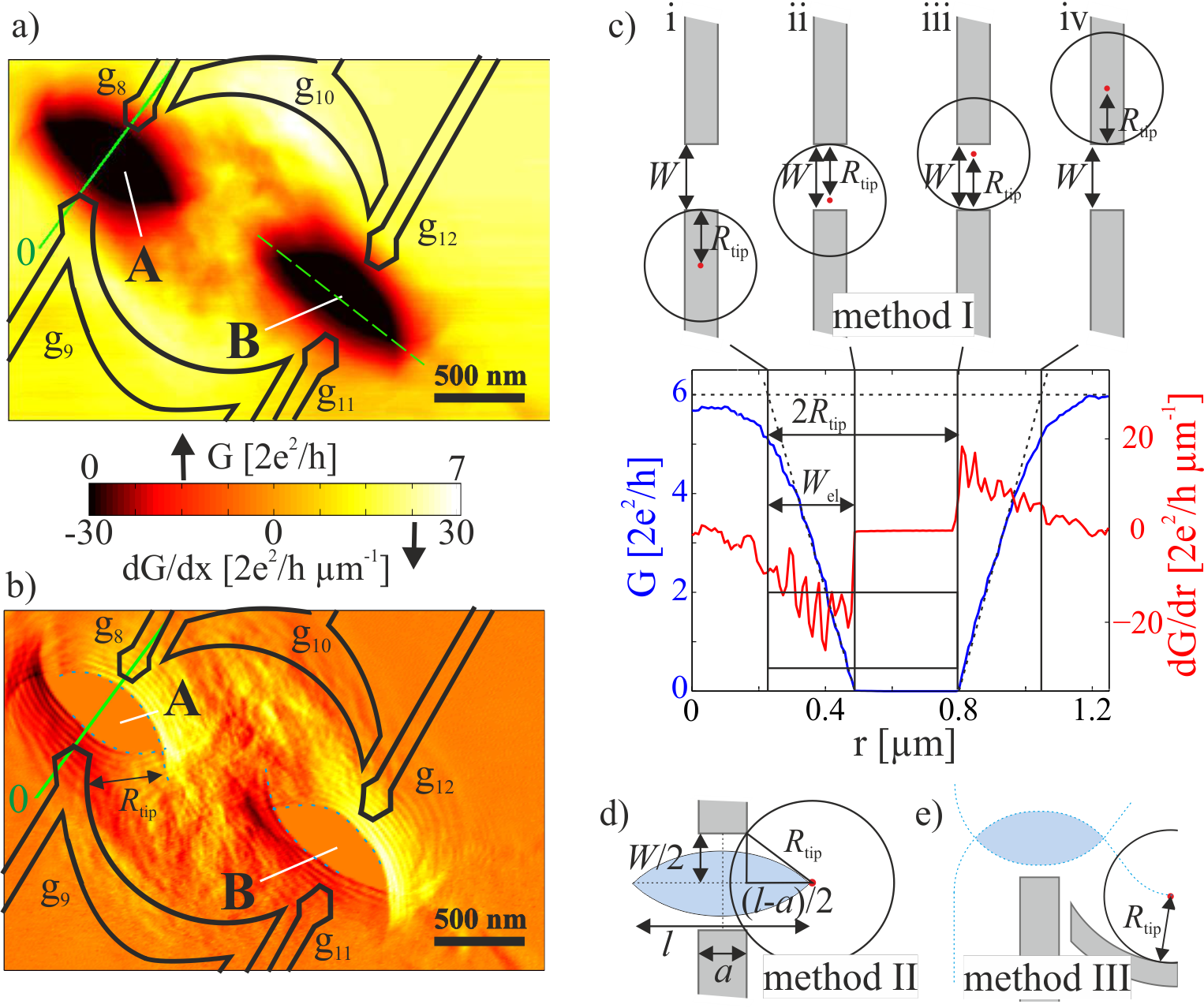} 
		\caption{a) Conductance map of scanning the tip ($V_\mathrm{tip} = -4$ V) above stadium II. The black lines indicate the outline of the biased top gates, grey ones the grounded gates, respectively. b) Numerical derivative $dG/dx$ of the conductance map. c) Line cut along the green line in a) and b) of the conductance (blue line) and the derivative (red line) with respect to the cut direction \textit{r}. The upper arrow indicates the diameter of the tip, the lower one the width of the constriction size $W_\mathrm{el}$. Above the line cut the tip position of the marked positions are sketched. The long-range action of the tip is negelected by extrapolating the approximative linear decrease in conductance towards the unperturbed QPC conductance value. d) Geometry of the length \textit{l} of the zero conductance region, the width $W_\mathrm{el}$ of the electronic constriction, and the gate width \textit{a}. e) The blue dotted line follows the path of the onset of current-flow between tip and gate.}\label{fig2}
\end{figure}

In \fref{fig2}a) the conductance $G$ of cavity II is shown as a function of tip position. The black lines correspond to the edges of the biased top gates ($V_{g8-12} = -0.4\,$V) which form the structure. The conductance decreases from approximately seven conductance quanta ($7 \times 2e^2/h$) for the tip at a position where it does not influence the cavity transmittance to zero in the vicinity of the two QPCs. The result are lens-shaped regions close to the two QPCs , labeled A and B in \fref{fig2}a) and b), similar as observed in \cite{Kozikov2013imaging}.

In order to understand how we can read the approximate size of the tip-depleted region from this image, we first concentrate on the solid green sectional line in \fref{fig2}a) and b). \Fref{fig2}c) shows the conductance and its derivative along this line together with schematic drawings of the tip position relative to the constriction. It is evident from the data and the schematics that $R_\mathrm{tip}\approx 0.28\,\mu $m. At the same time the electronic width of the constriction is seen to be $W_\mathrm{el} \approx 0.26\,\mu $m in agreement with the lithographic size and the depletion width caused by the applied voltage. This crude estimate, which we call method I in the following, regards the tip-depleted region to be hard-wall, simplifies the detailed geometry, and neglects all screening and stray-capacitance effects caused by the surface gates. It should therefore be taken as an order of magnitude estimate. The oscillations in the derivative of the conductance, also seen as fringes in \fref{fig2}b), reflect quantized conductance plateaux in the constriction formed between the tip and one of the QPC gates \cite{Kozikov2013imaging}.

\Fref{fig2}d) illustrates another geometric consideration for estimating $R_\mathrm{tip}$ from the extent of the lens-shaped region along the green dashed line (method II). One finds
\begin{equation*}
	R_{\mathrm{tip}} = \sqrt{[W_\mathrm{el}/2]^2 + [(l-a)/2]^2}\approx 0.33 \,\mu \textrm{m},
\label{Rlen}
\end{equation*}
where the width of the QPC gate is taken to be $a \approx 0.15 \,\mu \textrm{m}$, the extent of the lens-shaped region $l \approx 0.75 \,\mu \textrm{m}$, and the electronic width of the constriction $W_\mathrm{el} \approx 0.3 \,\mu \textrm{m}$. This result is in agreement with the previous estimate.

In \fref{fig2}b) we observe that the last fringe before depletion in the lens-shaped region can be followed into the interior as indicated by the blue dotted lines.  These lines run at approximately constant distance from the edge of the gate directly indicating $R_\mathrm{tip} \approx 0.5 \,\mu$m (method III) as illustrated in \fref{fig2}e). This estimate is an order of magnitude agreement with the previous ones, given the fact that the density in the cavity may be enhanced compared to the constrictions (although possibly reduced compared to the bulk), and given the distinct electrostatic environment formed by the surface gates.

\begin{figure}[h]
	\centering
	\includegraphics[scale=1]{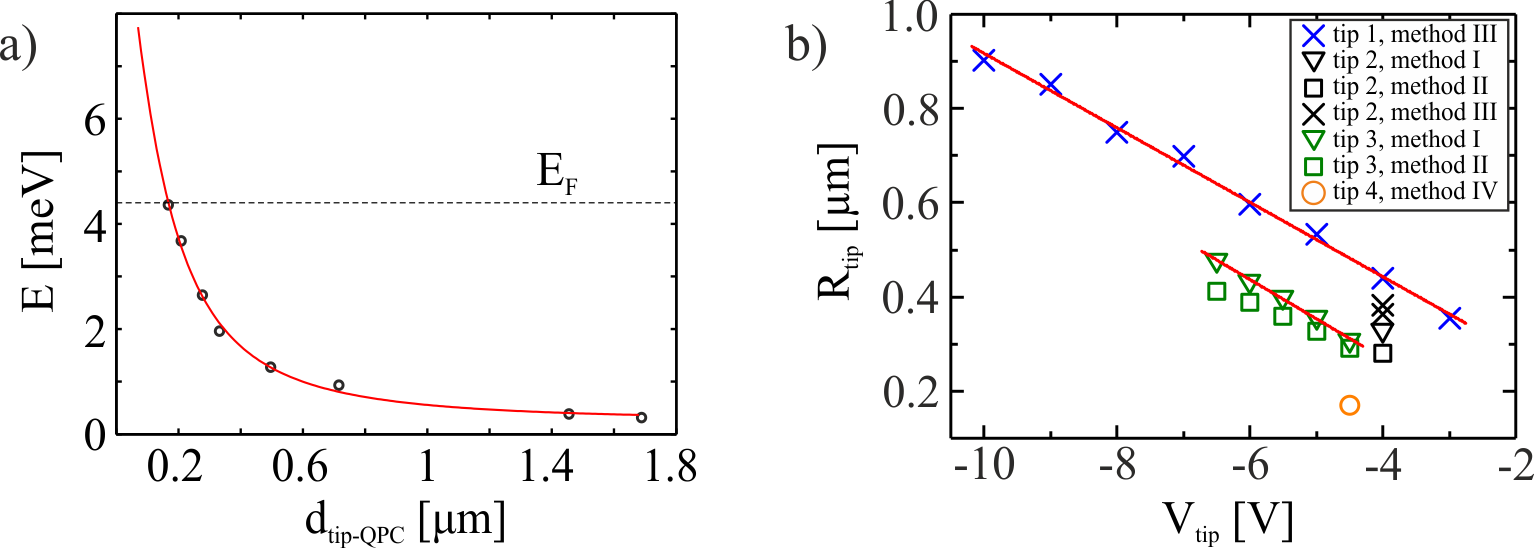} 
		\caption{a) Reconstruction of the tip induced potential (method IV), measured on the same sample as used in \cite{Kozikov2013imaging, Kozikov2013}, the tip-voltage is -4.5 V. The data points are converted from gate voltage to energy via a finite bias measurement and fitted with a Lorentzian. The dashed line indicates the Fermi energy level. b) Tip radii of four different tips, determined with different approaches shown in the text.}
		\label{fig3}
\end{figure}

All previous estimates of $R_\mathrm{tip}$ neglected the long-range tails of the tip-induced potential. The long-range capacitive coupling of the tip to a QPC \cite{Kozikov2013} can be used to determine this tail quantitatively. To this end the tip, kept at constant voltage, is placed at several positions along the transport axis of the QPC. At each point the QPC depletion gate-voltage is determined. Using finite-bias spectroscopy this gate-voltage can be calibrated to an energy scale \cite{rossler2011}. The resulting data for another tip than the previous, is shown in \fref{fig3}a), where the horizontal axis represents the distance from the tip to the center of the QPC.

We fit these data with a lorentzian line shape, since this was shown to be a reasonable approach \cite{Eriksson1996, Girard1993, Atlan1992, Pala2008, Pala2009, Chae2012, Martins2007, kivcin2005spatially, pioda2007discrete, Gildemeister2007}
\begin{equation*}
	E(x;E_0,A,x_0,\gamma) = E_0 + \frac{A}{(x-x_0)^2 + \gamma^2},
\end{equation*}
where $E_0$, $A$, $x_0$, and $\gamma$ are fitting parameters describing an energy offset, the peak amplitude, a position offset, and the line-width, respectively. The particular data shown in \fref{fig3}a) lead to $E_0 = (0.24 \pm 0.03)$ meV, $A = (0.372 \pm 0.005)$ meVnm$^2$, $x_0 = (-0.085 \pm 0.003)$ nm, and $\gamma = (0.160 \pm 0.005)$ nm. The intersection point of this reconstructed particular tip-induced potential with the Fermi energy of the electron gas gives an estimate of $R_\mathrm{tip} \approx (0.17 \pm 0.08) \,\mu$m (method IV). The largest contribution of the uncertainty of this estimate stems from the energy offset $E_0$, because this quantity results from the QPC gate-voltage to energy conversion. 

In \fref{fig3}b) $R_{\mathrm{tip}}$ of four different tips is shown as a function of tip-voltage. The tip-surface separation and the depth of the 2DEG are the same for all measurements shown (60 nm and 120 nm, respectively). The plot confirms that the different methods are consistent for a given tip. At a given tip-voltage different values of $R_\mathrm{tip}$ [compare the values of the tips used in figure 3a) and b)] are brought about by unintentional differences in tip fabrication and by modifications of the tip shape during topography scans \cite{Sellier2011}. The radius of the tip-depleted region is found to increase linearly with the tip voltage. This linear behavior is understandable since the Lorentzian is steep at the Fermi energy and thus can be approximated as a straight line within the given range of tip voltages.  For voltages above --3 V this assumption is not justified since the 2DEG is no longer depleted. The change of $R_\mathrm{tip}$ with $V_\mathrm{tip}$ is approximately 80 nm/V for tips of any radius given the tip-surface and surface-2DEG separation of 60 nm and 120 nm, respectively.

\section{Forming a cavity with the top gates}

\begin{figure}
	\centering
	\includegraphics[scale=1]{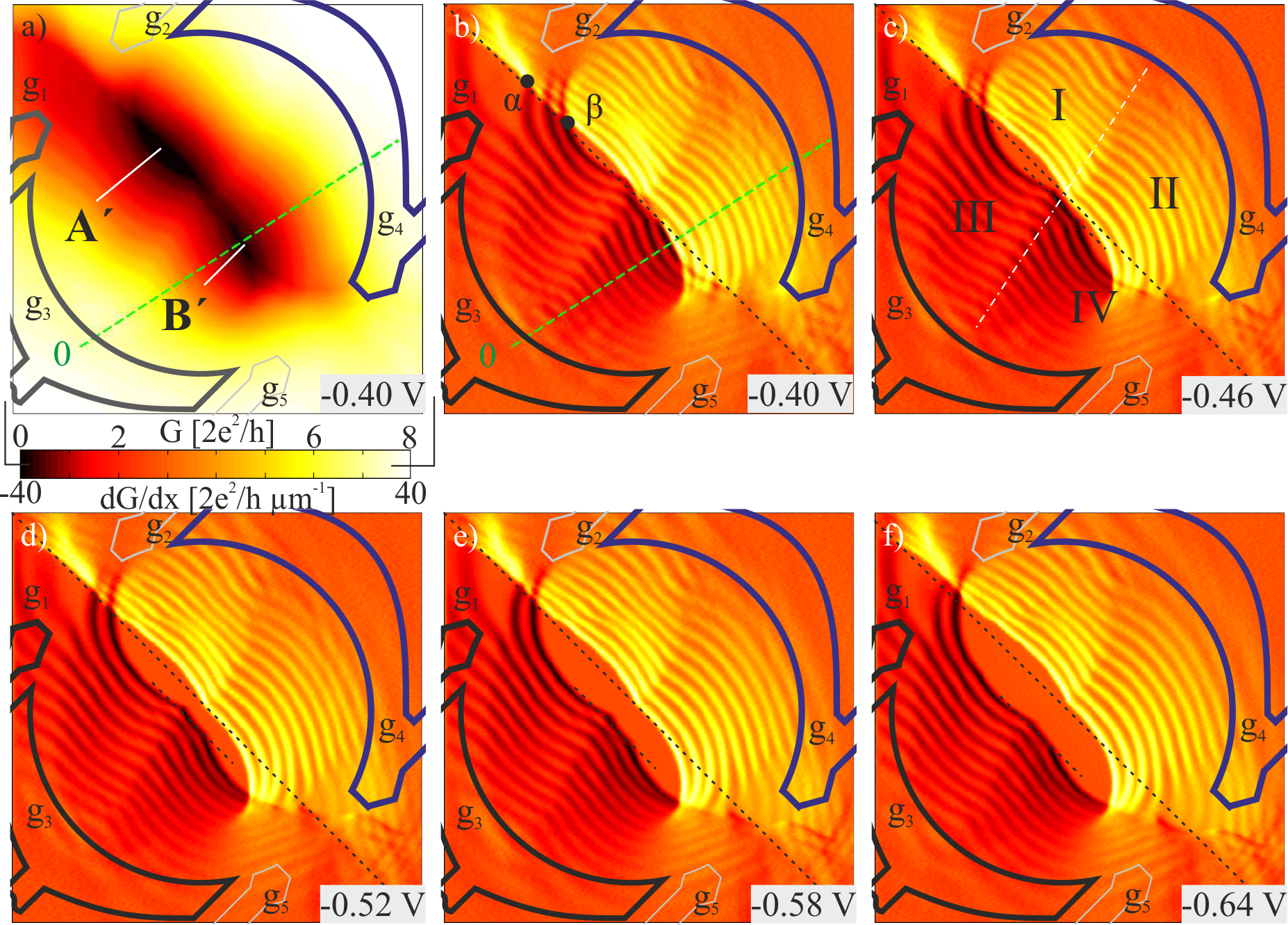} 
		\caption{Set of images taken with varying gate voltage applied to the upper top gate $g_4$ (solid purple line). The black solid line draws the biased gate with a fixed voltage throughout the whole set, the thin grey line belongs to grounded gates. a) Conductance map with $V_g = -0.4 \,$V. b)-f) Numerical derivatives with respect to x-direction. The black dashed lines gives a guide to the eye to follow the separation of the inner fringes.}
		\label{fig4}
\end{figure}

The tip characterized by the measurements of figure 2 is now used to find the change of the depletion width at the borders of the gate-defined cavity I [see \fref{fig1} a)] as a function of gate voltage. In order to get such spatial information, a set of 2d scans with the biased AFM tip and varying gate voltages is taken. For a first set of five scans the voltage on $g_4$ is varied while the voltage on $g_1$ and $g_3$ is kept constant at $-0.55\,$V. For the second set the roles of $g_4$ and $g_1$, $g_3$ are interchanged. The first set, shown in \fref{fig4}b)-f), leads to a fringe pattern in $dG/dx$ filling the whole cavity. In \fref{fig4}a) the conductance $G(x,y)$ corresponding to \fref{fig4}b) is given. The origin of the fringes is the same as in \fref{fig2}b): a quantized constriction forms between the tip-depleted region and one of the cavity gates. There are two groups of fringes, group I/II related to gate $g_1$ and $g_3$, and group III/IV related to $g_4$ [see labeling in \fref{fig4} c)]. The sequence of images in \fref{fig4}b)-f) shows that the group III/IV fringes shift in space with changing $V_{g4}$, whereas group I/II stays in place. This shift contains the desired quantitative information about the change of the depletion width.

\begin{figure}
	\centering
	\includegraphics[scale=1]{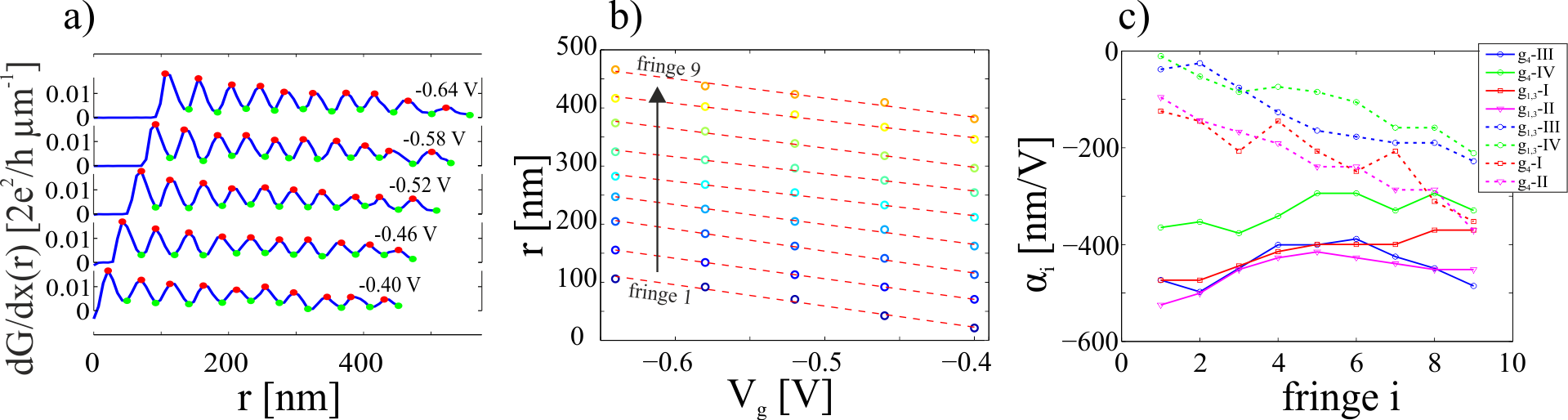} 
	\caption{a) Cut of the dG/dx maps of the set of varying $g_4$. The cut direction is the same as in \fref{fig4}a), but only for region IV, starting from the center. b) Position of the conductance plateaus (fringes in dG/dx) along the cut of the different gate voltages. c)All parameters characterizing the shift of the fringes of the different areas of both sets of measurements.}\label{fig5}
\end{figure}

The exact positions of the fringes can be extracted from the cuts [\fref{fig5}a)] along the green dashed line shown in \fref{fig4}b) for the five gate voltages applied to $g_4$. The fringes are labeled starting from the center of the cavity. These positions are indicated by filled circles in \fref{fig5}a). In \fref{fig5}b) we plot these points and fit them with a linear function of $V_g$ using
\begin{equation}
	l_i=\alpha_i \Delta V_g + l_{0,i},
\label{depl}
\end{equation}
where $\Delta V_\mathrm{g} = V_\mathrm{pinch-off}-V_\mathrm{g}$ is the difference of gate voltage from the gate pinch-off (-0.35 V), and $l_{0,i}$ is an arbitrary length offset irrelevant for the determination of the $\alpha_i$, with $i$ as the fringe number.

In \fref{fig5}c) we show the $\alpha_i$ determined from all scans. In addition, with the cavity divided into four regions I-IV [see \fref{fig4}c)], one characteristic cross-section is analyzed in each region for each scan. Points connected by solid lines refer to the situation where the constriction forms between the tip and the gate that is varied (case 1). Points connected by dashed lines refer to the situation where the gate is varied whose action on the constriction is screened by the tip (case 2). In the latter case the values of $\alpha_i$ are smaller, and they increase with fringe number (tip position) due to reduced screening of the gate voltage by the tip.
 The $\alpha_1$ parameter, which indicates the change of the depletion width with gate voltage, is of the same order of magnitude for all regions. At the same time the $\alpha_i$ vary only very little within each region in case 1.

\section{The origin of the fringe pattern shape}

The shape of the fringe pattern in figure \fref{fig4} does not reflect the cavity gate outline. Instead, it can be divided into four regions indicated in \fref{fig4}c). An additional effect not considered so far altering the tip-induced potential must be involved, since the tip-depleted region was found to be symmetric around the constrictions in section 3 [see figure 2a) and b)]. The fringes in these regions surround the lens-shaped regions A' and B' in \fref{fig4}a). This suggests that the constrictions involving the tip form mainly with the openings of the cavity, similar as discussed for \fref{fig2}. Additionally, the lens-shaped regions are shifted into the cavity from the geometric center of the constriction. A striking observation is made when the tip moves along the dotted line from point $\alpha$ to $\beta$ in \fref{fig4}b). While we would naively expect the conductance to increase we observe a decrease. Tentatively we ascribe this effect to enhanced screening of the tip-induced potential by the surface gates. By moving the tip closer to the constriction, its distance to the surface gates decreases, the tip-induced potential gets increasingly screened, and the 2DEG is no longer depleted below the tip.

\begin{figure}
	\centering
	\includegraphics[scale=1]{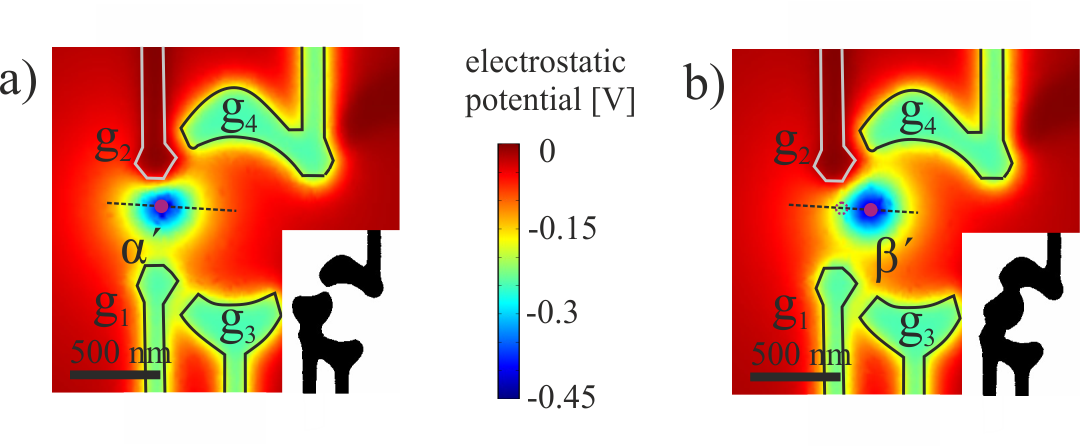} 
	\caption{Simulations of the electrostatic potential on the 2DEG induced by the biased tip. The gate configuration is the same as in the measurements above. The gate $g_2$ (grey) is grounded, $g_1$, $g_3$, and $g_4$ (black) are biased .  The insets show a 2d map at the Fermi energy, thus the depletion of the 2DEG. a) The tip is placed in the center of the entrance of the cavity. b) The tip is moved towards the cavity center until it blocks the constriction.}\label{fig6}
\end{figure}

Figure \ref{fig6} shows electrostatic simulations supporting this interpretation. Calculations were carried out with COMSOL \cite{Comsol} treating the 2DEG as a grounded metallic plane 100 nm below the metallic top gates. The GaAs material was modeled as a dielectric with $\epsilon = 13$. The tip, implemented as a metallic cone with a hemisphere with radius 50 nm at its end, is placed 70 nm above the surface. We determine the induced density in the 2DEG and consider regions to be depleted if the induced density exceeds the sheet density of the electron gas.

 In \fref{fig6}a) the tip is placed in the opening of the cavity [position $\alpha '$ in \fref{fig6}a)]. The screening of the induced potential by the gate $g_2$ leads to an open conductance channel, as indicated in the inset showing the depletion area below the gates and the tip. In \fref{fig6}b) the tip is moved along the dotted line towards the cavity center until the constriction is closed (position $\beta '$). These simulations show that the zero conductance regions A' and B' are shifted relative to the constriction center into the cavity in the presence of grounded gates close to the constriction.

\section{Conclusion}

We have presented methods for estimating the size of the tip-induced depletion region in the 2DEG using a biased AFM tip and the investigation of the shape of a ballistic cavity. Even though most of the methods use simplified geometric assumptions their errors may play a minor role compared to electrostatic screening effects encountered in the experiments. But even with such limitations, fully quantized transport resulting in an unprecedented clear fringe pattern covering the entire stadium is observed. The findings are pointing towards the accessibility of the local density of the electronic states in ballistic cavities for optimized structures regarding tip potential screening by the top gates.

\section{Acknowledgements}
We acknowledge financial support from the Swiss National Science Foundation, the
NCCR ”Quantum science and Technology” and ETH Z\"{u}rich.

\end{document}